\documentstyle[epsf]{aipproc}

\def\anti{\overline}
\def\br{B}
\def\fbi{~{\rm fb}^{-1}}
\begin{document}
\hspace*{3.7in}{\bf IUHET-421}\\
\hspace*{3.85in}{\bf March 2000}
\title{SUSY Thresholds at a Muon Collider\thanks{To appear in the 
{\it Proceedings of the 5th International Conference on Physics Potential and
Development of $\mu^+\mu^-$ Colliders (MuMu99), San Francisco, CA, 15-17 
December 1999.}}}

\author{M. S. Berger}

\address{Physics Department, Indiana University, 
Bloomington, IN 47405}

\maketitle

\begin{abstract}
One of the useful features of muon colliders is the naturally narrow spread
in beam energies. Measurements of threshold cross sections then become a 
prime candidate for precision measurements of particle masses, widths, and 
couplings as well as determining particle spin. We describe the potential 
for measuring cross sections near threshold in supersymmetric theories.
\end{abstract}

\section*{Introduction}
%
%
%
%

Muon colliders have negligible bremstrahlung and the beam energy can be 
measured very accurately by muon precession in the collider ring. Momentum
spreads as low as $\Delta P/P=0.003\%$ are thought to be achievable for 
a low-energy collider\cite{muon,expression}. 
The beam energy could be determined with a precision 
of $\Delta E/E=10^{-6}$ by measuring the time-dependent decay asymmetry
resulting from the naturally polarized muons\cite{rt}. In addition, 
initial state radiation is smaller than at a electron-positron machine.
These features make muon colliders especially useful for studying narrow 
resonances\cite{higgs}, and for measuring production
cross sections near threshold where they change very 
rapidly\cite{iowa,fermi2,cernreport}. However the cross sections are smaller
near threshold, so high luminosity is required.

An important feature of cross sections near threshold is that they isolate
the effects of the width and spin of the produced particle.
The threshold cross section is a function of the particle's mass, decay 
width, spin, and coupling strength(s), and it (largely) factorizes 
near the threshold into an overall normalization and an energy-dependent part
(profile). 

The energy profile depends on the particle mass which dictates where in energy
the cross section begins to rise; we say the cross section ``turns on'' near
$2m$. The particle width and spin govern how fast
the cross section rises with energy. The width is important because, on the 
one hand, we can think of the particle being produced and then subsequently 
decaying (the narrow width approximation); but on the other hand, one can 
more accurately view 
the process as one in which the decay products are the final state
and the decaying particle is a virtual particle. The rate the cross section
rises depends then on the required ``off-shellness'' of this virtual particle.

As an example consider
the cross section for $\mu^+\mu^-\to W^+W^-$ followed by the decay 
$W\to f\overline{f}^\prime$ is given by\cite{mnw}
\begin{eqnarray}
&&\sigma(s)=B_{f_1\overline{f}_2}B_{f_3\overline{f}_4}\int _0^s\:ds_1\rho(s_1)
\int _0^{(\sqrt{s}-\sqrt{s_1})^2}\:ds_2\rho(s_2)\sigma (s,s_1,s_2)\;,
\end{eqnarray}
where $\sigma(s,s_1,s_2)$ is the cross section for producing two virtural 
$W$ bosons with invariant masses-squared of $s_1$ and $s_2$, and 
$B_{f_1\overline{f}_2}$ is the branching ratio for the decay 
$W\to  f_1\overline{f}_2$. The weighting factor $\rho(s)$ arises from the 
$W$ boson propagator and controls the rise of the cross 
section,

\begin{eqnarray}
&&\rho(s)={1\over \pi}{{\sqrt{s}\Gamma(s)}\over 
{(s-M_W^2)^2+M_W^2\Gamma (s)^2}}
\;.
\end{eqnarray}

The cross section $\sigma(s,s_1,s_2)$ is controlled by kinematics and angular
momentum conservation for the region that $\rho(s_1)$ and $\rho(s_2)$ have 
significant support near $s_1,s_2\approx M_W$.
The couplings and radiative corrections are approximately
energy-independent over this narrow range, and do not impact on the energy 
profile, but do
affect the overall number of events expected.
  
\section*{Chargino Production}

\begin{figure}[htb]
\leavevmode
\begin{center}
\epsfxsize=3.5in\hspace{0in}\epsffile{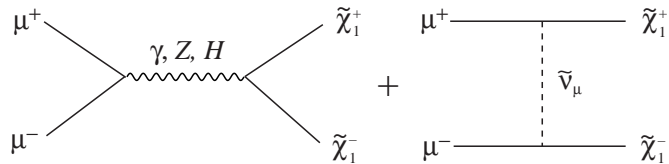}
\end{center}
\caption[]{\footnotesize\sf Feynman diagrams for chargino pair production.}
\label{feynman}
\end{figure}

Supersymmetric charginos can be pair produced at a muon collider.
The two contributing Feynman diagrams for 
$\mu^+\mu^- \to \tilde{\chi}_1^+\tilde{\chi}_1^-$ is shown in 
Fig.~\ref{feynman}.
These two diagrams interfere desctructively over most of parameter space.
The threshold measurement of the process 
$\mu^+\mu^-\to \tilde{\chi}_1^+\tilde{\chi}_1^-$ yields the following:
a measurement of
the $\tilde{\chi}_1^\pm$ mass, an indirect measurement of the 
$\tilde{\nu}_\mu$ mass, and a measurement of the overall normalization of 
the cross section.

The supersymmetric masses are known functions of the undetermined fundamental 
parameters of a given supersymmetric model, and are a window on the 
details of the supersymmetry breaking. For example the chargino mass matrix
is 
\begin{eqnarray}
M_{\tilde{\chi}^\pm}=\left( \begin{array}{c@{\quad}c}
M_2 & \sqrt{2}M_W\sin \beta \\
\sqrt{2}M_W\cos \beta & -\mu
\end{array} \right)\nonumber
\end{eqnarray}

The fundamental parameters of the supersymmetric model are 
the $SU(2)$ gaugino mass $M_2$, the 
supersymmetric Higgs mass $\mu$, and the ratio of the vevs of the two
Higgs doublets' vacuum expectation values, $\tan \beta = v_2/v_1$.
The physically measured chargino masses $M_{\tilde{\chi}^\pm_1}$ and 
$M_{\tilde{\chi}^\pm_1}$ are obtained by diagonalizing this matrix. 
The masses and widths are complicated 
functions of fundamental parameters, but accurate measurements 
of the chargino masses constrain one
to be on a slice through the parameter space of fundamental parameters.

\begin{figure}[htb]
\leavevmode
\begin{center}
\epsfxsize=3.25in\hspace{0in}\epsffile{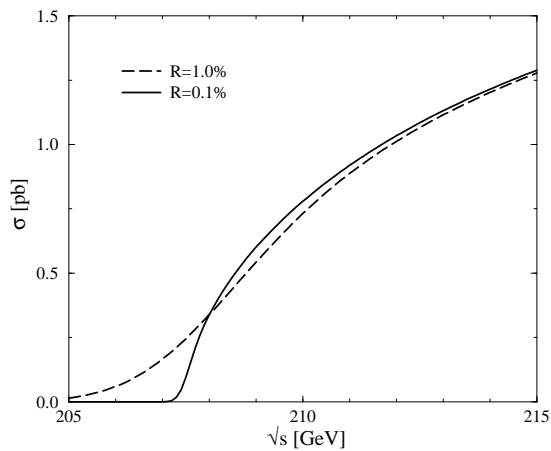}
\end{center}
\caption[]{\footnotesize\sf A muon beam with a narrow beam spread ($R=0.1\%$)
preserves the rapid rise in the chargino pair production threshold cross 
section.  The threshold cross section convoluted with a beam with $R=1.0\%$ is 
shown for comparison. The chargino parameters are $M_2=100$~GeV, 
$\tan \beta =4$, and $\mu>>M_2$.}
\label{beamsm}
\end{figure}

The impact of different amounts of beam smearing is shown in Fig.~\ref{beamsm}.
The narrow beam spread ($R=0.1\%$, where $R$ is the rms spread of the energy 
of a muon beam) of a muon collider can prove especially 
advantageous when the widths of the produced particles is substantially less
than a GeV.
  
\begin{figure}[htb]
\leavevmode
\begin{center}
\epsfxsize=3.25in\hspace{0in}\epsffile{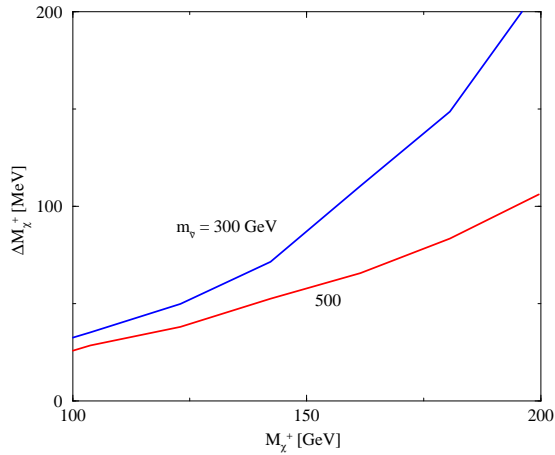}
\end{center}
\caption[]{\footnotesize\sf The $1\sigma$ precision obtainable in the 
chargino mass taking $m_{\tilde{\nu}} = 300$ and $500$~GeV. The precision is
better for {\it larger} sneutrino mass because the contribution from the
$t$-channel sneutrino exchange diagram destructively interferes with
the $s$-channel diagrams.}
\label{prec}
\end{figure}

The chargino production cross section decreases with increasing chargino mass.
Therefore the precision with which the mass can be measured is better at 
smaller values of the mass. Figure~\ref{prec} shows the expect precision with 
100~fb$^{-1}$ integrated luminosity for sneutrino masses of 300 and 500~GeV. 
For a lighter sneutrino, for which the destructive interference between
the $s$-channel and $t$-channel graphs is more severe, the precision obtained
is reduced.
The sneutrino mass can be measured to about 6~GeV accuracy for 
$m_{\tilde{\nu}} = 300$~GeV and to about 20~GeV accuracy for 
$m_{\tilde{\nu}} = 500$~GeV\cite{bbh,fermi1}. 

\section*{Top squark production}

Scalar production is $p$-wave suppressed, so the threshold production cross
section looks very different than for fermion pair production. For example one
can compare top quark production to top squark production near 
threshold\cite{berger}. The top squark cross section shows a 
slow $\beta ^3 $ rise with little resonance structure\cite{mod}.
The top squark threshold profile is very different that of the top quark.
By measuring energy profile, one can determine the produced particle's
spin.
If the squarks are for some reason especially long-lived, 
then bounds states might be formed
that would appear in the threshold cross section\cite{cernreport}.

\section*{Threshold Higgs Production}

An accurate determination of the supersymmetric Higgs boson mass $m_H$ can be 
obtained by measuring the threshold cross section for the Bjorken 
Higgs-strahlung process\cite{bj} $\mu^+\mu^-\to ZH$. The Higgs boson could
be discovered in the $ZH$ production mode by running the collider well above 
threshold. For $m_H<2M_W$, as one expects in supersymmetric models, the 
primary decay channel is $b\overline{b}$, and most backgrounds can be 
eliminated by $b$-tagging.

\begin{figure}[htb]
\leavevmode
\begin{center}
\epsfxsize=5.0in\hspace{0in}\epsffile{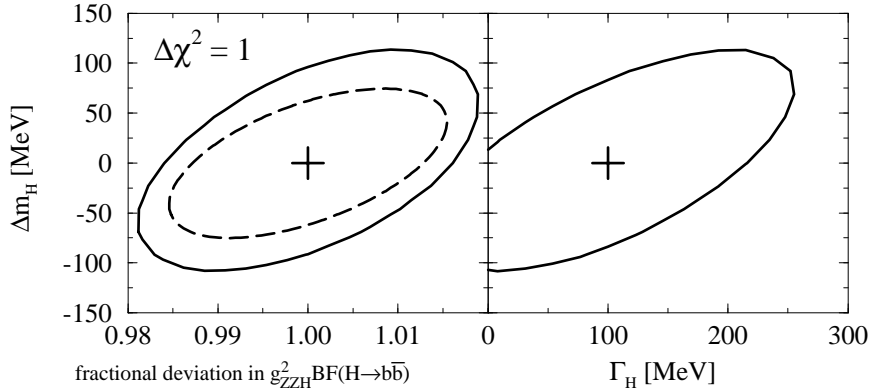}
\end{center}
\caption[]{\footnotesize\sf A simulatneous measurement of the Higgs mass and
the overall rate $g_{ZZH}^2\br(H\to b\anti b)$ s shown on the left. A 
measurement of the Higgs width is shown on the right.}
\label{zhsin2}
\end{figure}

The Higgs cross section near threshold is sensitive to the Higgs mass and 
width. In Fig.~\ref{zhsin2} the results\cite{bbgh,fermi3} 
for a Higgs mass around 100~GeV are
displayed. The solid curves show the $\Delta \chi ^2=1$ contours
for determining
the Higgs mass versus $g_{ZZH}^2\br(H\to b\anti b)$, or versus
$\Gamma_H^{}$, by devoting
$100/3\fbi$ to each of the c.m.\ energies
$\sqrt{s}=M_Z+m_H+0.5$~GeV, $\sqrt{s}=M_Z+m_H+20$~GeV
and $\sqrt{s}=M_Z+m_H-2$~GeV at a muon collider.
included. The dashed curve shows the
$\Delta\chi^2=1$ contour that results when $\Gamma_H$ is negligibly small and
50~fb$^{-1}$ is devoted to each of the c.m.\ energies $\sqrt s = m_H + M_Z +
0.5$~GeV and $\sqrt s = m_H + M_Z  + 20$~GeV.

\section*{Conclusions}
   
Muon colliders offer advantages for performing threshold cross 
section measurements: no beamstrahlung, small beam energy spread, reduced 
initial state radiation.

Precise measurements of particle masses and widths are 
possible, and the shape of the threshold cross section
can be measured to determined the produced particle's 
spin.

\section*{Acknowledgement}
Work supported in part by the U.S. Department of
Energy under Grant No. DE-FG02-95ER40661.

\end{document}